\documentclass{article}
\usepackage[utf8]{inputenc}
\usepackage{authblk}
\usepackage{setspace}
\usepackage[margin=1.25in]{geometry}
\usepackage{graphicx}
\graphicspath{ {./figures/} }
\usepackage{subcaption}
\usepackage{algorithm}
\usepackage{algorithmic}
\usepackage[utf8]{inputenc}
\usepackage{graphicx}
\usepackage[T1]{fontenc}
\usepackage{amsmath}
\usepackage{multirow}
\usepackage{booktabs}
\usepackage{cuted}  
\usepackage{caption}     
\usepackage{csquotes}    
\usepackage{algorithm}
\usepackage{algorithmic}
\usepackage{amssymb}
\usepackage[english]{babel}          
\usepackage{microtype}
\usepackage{geometry}
\usepackage{setspace}
\usepackage{lmodern}
\usepackage{courier}

\usepackage{xcolor}
\definecolor{primarycolor}{RGB}{0, 51, 102}
\usepackage[
colorlinks=true,
linkcolor=primarycolor,
urlcolor=primarycolor,
citecolor=primarycolor,
pdfborder={0 0 0}
]{hyperref}

\usepackage{fancyhdr}
\usepackage{amsmath}

\usepackage[style=nejm, 
citestyle=numeric-comp,
sorting=none,
backend=bibtex]{biblatex}
\title{GCTAuto-encoder: A Cross‑modal Framework for Security Flaw Detection in IoT Networks}

\author{Najmieh Sadat Safarabadi}

\makeatletter
\renewcommand{\@makefnmark}{}
\makeatother

\affil[]{Department of Computational Linguistics, University of Stuttgart, Stuttgart, Germany.}

\date{}

\begin{document}

	\maketitle
	
	\begin{abstract}
		IoT encompasses a wide range of physical entities, from smart home devices to autonomous vehicles, creating a complex environment that employs discrete security models to tackle its diversity. Due to the heterogeneous nature of the IoT infrastructure, the sub-systems are often vulnerable to various network attacks. To address these ongoing limitations, modern security systems must be more credible to ensure security and privacy for IoT applications. Moreover, a highly secure IoT system demands real-time insight, which calls for insight collection from decentralization of the edge of the computing layer. At the same time, this diversity of the IoT ecosystem entails a unified security model that is applied at the foundational level. One of the counterparts to this requirement is edge intelligence, which is considered the direct approach to the diversity of the involved devices. One of the practical goals of edge intelligence in the IoT ecosystem is to extract insight from local data; then the security model leverages the data gathered from this stage to build local node protections, and its integration with AI models can present an advanced security solution to this challenge. In this direction, the current research proposes a novel deep learning algorithm utile for effective intrusions to detections on the edge of intelligence, supported by a cloud-based IoT framework. This study evaluates the proposed cross‑modal deep learning algorithm and compares its performance against baseline models. Accordingly, the contribution of the current research is a cross‑domain Deep Neural Network (DNN) algorithm that facilitates intrusion detection.The objective is to assess a multi‑method deep learning model to detect intrusion in IoT systems on the edge of intelligence based on detection of communities with modelled attention. In this context, the study evaluates GCTauto-encoder, a novel framework that integrates edge intelligence within a comprehensive system to identify security flaws. The proposed model demonstrates that it significantly improves performance and efficiency. Evaluated on a network intrusion IoT dataset covering different network intrusion scenarios, it achieved a 0.908 accuracy improvement over benchmark models, reduced learning loss to 0.00156, and outperformed existing approaches within the evaluated setting.
	\end{abstract}
	
	
	\section{Introduction}
	Internet of Things (IoT), expanding due to its capacity for integration with artificial Intelligence, has expanded the application of IoT to machinery data, data-driven techniques, modern networks, mutual services across industries, and is promoting further expansion \cite{Fortunato2010}. IoT has notable applications across smart healthcare, smart home, smart transportation, and smart industry—with fine-grained practices seen in areas like localization, predictive monitoring, autonomous driving, and intelligent control. A robust IoT ecosystem must establish credibility through strong security and privacy measures. Given the diverse network architecture of IoT with numerous nodes, these systems are vulnerable to various cyber-attacks, especially due to the massive data flows they generate over time. To protect IoT applications, traditional security measures must be strengthened to address system-specific vulnerabilities of IoT \cite{Ring2019}. The IoT's vast network has a dynamic nature where intrusion risks necessitate robust security frameworks, especially in sensitive areas like healthcare, where data privacy is crucial. In this regard, edge computing and artificial intelligence in the paradigm of edge of intelligence play crucial roles in detecting network vulnerabilities, enabling real-time decision-making and efficient communication between devices. The availability of blockchain technologies aligns with modern intrusion detection systems (IDS) and improves security in IoT network infrastructures. On the other hand, methods based on Edge Artificial Intelligence (EdgeAI) demonstrate notable effectiveness in identifying network patterns, thereby mitigating tampering risks in traditional IDS. Motivated by these observations, the remainder of this paper is structured as follows. Section two reviews the existing literature on the IoT ecosystem, with a focus on security, privacy, and the role of AI Models. section three to seven describes the research methodology adopted in this study. Section eight presents and discusses the key findings. Finally, section nine to ten concludes the paper, summarizing the main insights and outlining directions for future research.
	
	\section{Literature Review}
	\noindent The Internet of Things is no longer in perspective, and it has a key constituent in the billions of communication units. The far-reaching targets of closing the gap between physical and virtual entities are in close proximity in the current area. This opens numerous gates for integrations of artificial intelligence, while its production of massive data circulations imposes a demand for real-time data analysis. As IoT ecosystems have grown in complexity, security models have shifted accordingly. Earlier approaches relied on unified, one-size-fits-all mechanisms applied at the foundational level. However, as the diversity of devices, from healthcare systems to autonomous vehicles, expanded, the limitations of such uniform models became apparent. A growing body of research work has moved toward discrete security models that operate locally, extracting insight from device-level data and progressively analyzing them as a whole \cite{Wang2025EdgeDataAnalysis}. This view is supported by studies demonstrating that direct, localized approaches are better suited to heterogeneous network security architectures \cite{Al-Fuqaha2015IoTSurvey}. Proceeding literature has concentrated on methods to secure IoT wireless network paradigms and expand its high capacity even further, leveraging  Artificial intelligence. Based on this, integration of machine learning models can be used as catalysts for fast detection of security flaws. This provides a systematic understanding of better approaches revolves around three broad themes that emerge from this literature on modern AI security models \cite{Sicari2015IoTSecurityPrivacy}: (i) device-level defence with system-wide synthesis, (ii) integration of adaptive AI techniques, and (iii) the privacy–functionality trade-off in large-scale networks. Moving forward, recent research suggests the establishment of clusters of controllers, each responsible for specific decision-making tasks—a concept that maps naturally onto AI, where each controller functions as an intelligent agent processing sensor data from its sub-ecosystem. To maintain real-time security across the system, a dynamic slave controller can be assigned within each cluster of nodes. This continuously assesses task allocation, circulation, and migration as network conditions change \cite{NewmanGirvan2004}. An instance of such a network modelling is HetNets \cite{Lin2017IoTSurvey}, whose functionality relies on seamless coordination between different tiers of base stations, while their security is challenged by the expanded attack surface introduced by densely distributed access points \cite{Raghavanetal2007}. Such a network presents multiple security challenges, including the need for accurate identification of threats across the system \cite{SubeljBajec2011a}, \cite{Blondeletal2008}. These concerns, though exemplified in specific network architectures, are part of a wider set of challenges driving the field. Furthermore, the trajectory of research in this area reflects a move toward an intelligent model that is capable of more precise detection and classification of security events across the system \cite{SobolevskyCampari2014}. 
	On the edge of intelligence, one of the most important procedures is the selection of an optimum frameworks or models to further harvest the security mechanism and to implement the most suitable artificial intelligence solutions to investigate the underlying patterns in the sensors' collected data. In this domain, the earliest AI models applied to network security were rule-based expert systems \cite{Vehlowetal2013}, which encoded human knowledge to detect known attack patterns \cite{Gregory2011, Zhangetal2007}.
	As IoT ecosystems grew in complexity, however, a new approach became necessary: a large-scale entity that cooperates through localization-based methods to achieve system completion. These were followed by shallow machine learning classifiers, such as decision trees, support vector machines, and naive Bayes trained on labelled traffic data for intrusion detection. Neural networks with a single hidden layer later enabled more flexible pattern recognition, including anomaly detection  \cite{Sperotto2010} beyond predefined signatures. Despite their limitations, these early models laid the foundation for today's deep learning approaches by demonstrating that automated, data-driven methods could identify malicious activity without explicit programming. A later transfer-based approach for the same fault was employed in \cite{Zhao2019TransferFaultDiagnosis}. In earlier attack detection and fault‑diagnosis pipelines, the Wavelet transform constitutes the primary preprocessing step \cite{Yang2003WaveletFaultDiagnosis}, extracting time–frequency representations from raw vibration signals. These representations are subsequently classified using soft-max regression, which provides the baseline linear model in the overall framework. Finally, variational and stacked auto-encoders (SAE) enable unsupervised feature learning through stacked reconstruction, while DNN advances further into fully end-to-end supervised deep learning, together capturing the progression from shallow signal processing to modern deep learning for bearing fault diagnosis. In a later progressive solution, a transfer learning-based approach was investigated. Additionally, in cases where a larger amount of data processing was required, Bayesian \cite{Psorakisetal2011}, Weibull distribution, and deep belief network (DBN) were applied to evaluate the bearing degradations \cite{Zhao2019DeepLearningIIoT}, \cite{Zhang2016WeibullDBN}. In short, the literature on IoT ecosystems has evolved from unified security models toward localized, device-specific approaches, driven by the growing diversity of devices and data flows. However, existing studies have typically implemented security controls and data‑analytics pipelines in isolation, resulting in systems where intrusion detection, encryption, or access control are designed without regard to how diagnostic or predictive models consume the same operational data \cite{Lin2017IoTSurvey}. This gap motivates the present study, which investigates how graph modelling of deep learning approaches equipped with contrastive learning can bridge security and local and global data analysis in heterogeneous IoT network environments and their protection systems.
	\section{Preliminaries} 
	\subsection{The Edge Of Intelligence}
	The IoT paradigm builds upon several established technologies that collectively enable real-time coordination at scale. Emerging technologies like the Internet of Vehicles,  Internet of Medical Things, Industry 5.0, Industrial IoT, cloud computing, and the Internet of Everything require highly coordinated management capabilities. One primary goal of these established technologies is to enable real-time machine-to-machine decision-making. To achieve optimal outcomes in distributed AI inference at the edge, scalability and reliability are critical. This mainly because edge deployments must handle variable workloads across heterogeneous devices while maintaining consistent inference quality. In this regard, an effective load balancing scheme, such as ESCALB, a slave controller allocation-based approach, is essential for multi-domain SDN-enabled IoT networks. Software-Defined Networking (SDN) \cite{kreutz2015sdn} can support these requirements by enabling adaptive traffic routing between edge nodes, dynamic resource allocation based on inference demand, and centralized visibility into distributed compute health. SDN's programmable control plane allows the management system to orchestrate inference tasks across the edge fabric, redirecting workloads when nodes fail or become congested.
	\begin{figure*}[h]
		\centering
		\includegraphics[width=0.9\textwidth]{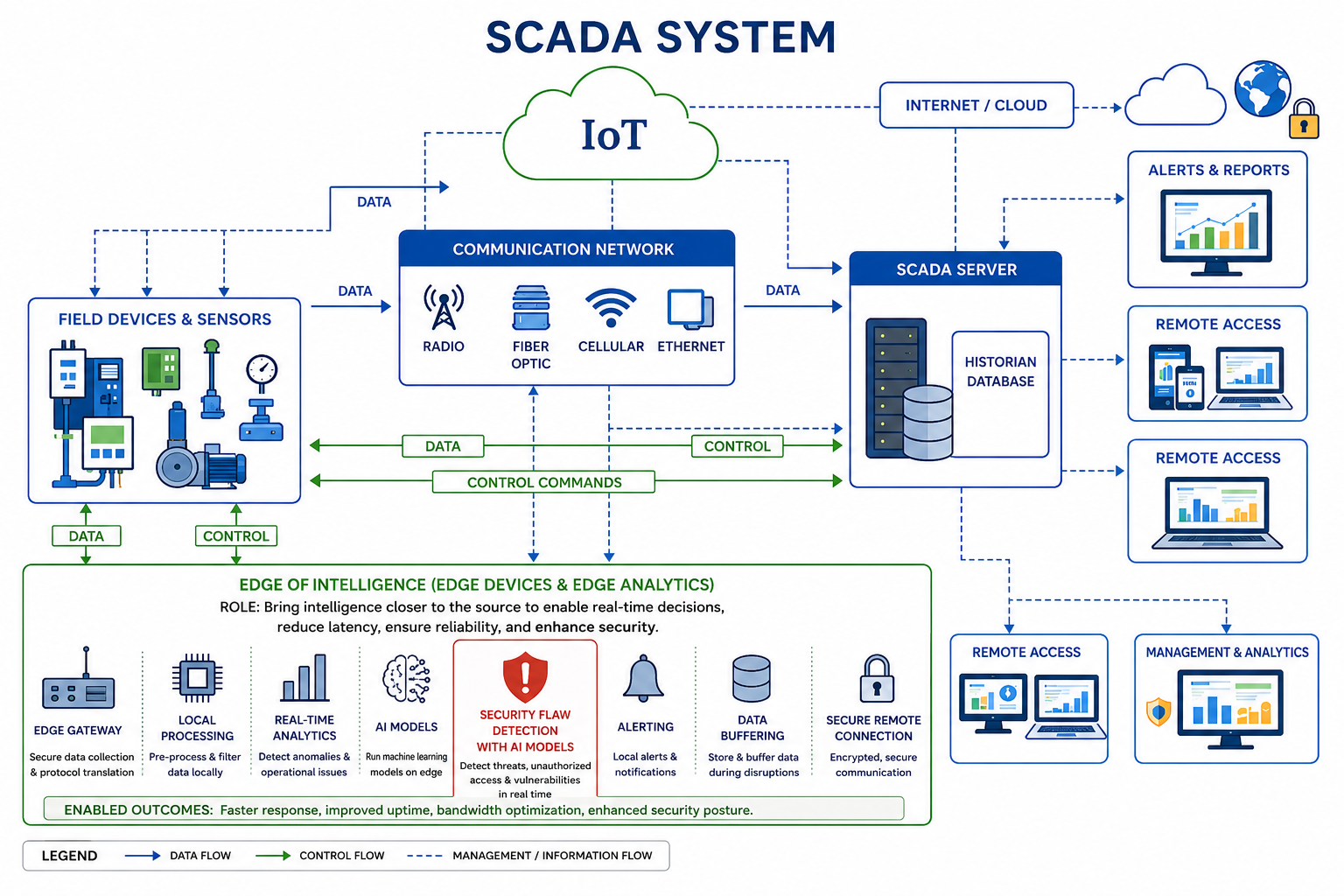}
		\caption{Supervisory Control and Data Acquisition in IoT infrastructure }
		\label{FIG:SCADA}
	\end{figure*}
	
	\subsection{The SCADA System}
	The behavioural insights regarding physical entities, IoT gateways, data flow, and control mechanisms are provided by SCADA systems \cite{boyer2009scada} at each edge of intelligence. SCADA systems play a key role since they directly interact with and collect data from devices like sensors and motors. This serves as both event recorders and decision-makers. They can communicate issues within the system, which helps to reduce downtime. Current SCADA systems can greatly benefit from IoT integrations. This integration enhances features such as improved interpretability and flexible scalability. Based on the \ref{FIG:SCADA} image, the SCADA system brings intelligence closer to the source. The diversity within the IoT sub-ecosystem, combined with the connectivity of all components, can generate significant network traffic. The complexity of computing in IoT increases due to its dispersed nature. A centralised controller managed by SDN (Software-Defined Networking) can provide a global perspective. Such a modelling can lead to improved performance and more coherent decision-making across the distributed environment. The SDN controller sees the entire IoT network topology. This makes anomaly detection \cite{Sperotto2010} and troubleshooting far easier than with isolated switches. For instance, the control plane can act as a decision-maker during security incidents, managing overloaded traffic, data preservation, and authentication. This significantly enhances the efficiency of the IoT system. The SDN controller can be programmed to dynamically adjust network behaviour based on SCADA events (e.g., if a critical alarm triggers, prioritise that substation's traffic), and Northbound APIs expose network state to SCADA security monitoring tools. Moreover, the quality of service (QoS) based on the SCADA protocols (Modbus, DNP3, IEC 61850) is a critical consideration for achieving optimal results in IoT networks \cite{galloway2013scada}. With the edge of intelligence, one of the primary objectives is to maintain high-quality service (QoS); therefore improving communication metrics is key for overall performance. In SDN, a dynamic slave controller is utilized within a multi-criteria decision-making strategy to enhance decision-making by selecting effective controllers based on varying data flows and resource availability. Additionally, to efficiently identify network vulnerabilities, Edge Intelligence (EI) and Edge Artificial Intelligence (EdgeAI) \cite{shi2020edgeai} have emerged as promising solutions within the IoT ecosystem that can be easily integrated with modern AI models. By bridging the gap between traditional IoT security infrastructure \cite{sicari2015iotsecurity} and intelligent automation, they open new avenues for efficiency and adaptability. 
	\section{Problem Definition}
	While the above approaches and technologies enable high capability of IoT systems in modern networks, they still face a demand for effective decision-making in relation to IoT security vulnerabilities. A key limitation of conventional IoT networks is that each node operates based on only local information \cite{gelenbe2016iot}. While individual nodes may possess valuable insights about their immediate environment, there has not been systematic research work to consolidate these fragmented perspectives \cite{yang2019iotreview} \cite{yang2017iotsecurity} into a coherent global view. Accordingly, the following problem arises when the system does not benefit from both local responsiveness and centralized aggregation. The problem becomes more pronounced when the typical limitations of a centralized computing architecture lack contextual knowledge \cite{zhang2021context}, delays in centralized reporting, alarm parallelization, and inadequate real-time notifications. These issues are considered security flaws in this structure. If the focus is on a centralized computing architecture for IoT, it enforces restrictions on IoT platforms, particularly when different distributed components are involved. The aim is to pass the detection to a system that first assess on local and then leverage the knowledge as a whole. A centrally placed model that receives inputs from all local nodes can effectively act as the aggregation point. In this manner, the approach benefits from both centralized and decentralized approaches. Conventional security systems have limited credibility to provide a solid foundation for security and privacy in IoT applications. On the other hand, the Intrusion Detection Systems (IDS) and Intrusion Prevention Systems (IPS) \cite{scarfone2007guide} must be designed in a way that is intelligent and distributed within modern cloud-based IoT platforms. This means that the components of the IDS and IPS can vary across different sections of the platform. The goal is to transition to a centralized architecture for the controller to streamline operations. This becomes increasingly critical as variation in IoT scenarios and requirements grows. In the IoT ecosystem, Distributed Intrusion Detection Systems (IDS) equipped with Edge Intelligence (EI) \cite{shi2019edgeintelligence} are essential for protection against Distributed Denial of Service (DDoS) attacks and their variations. Moreover, making IDS and Intrusion Prevention Systems (IPS) autonomous in their decision-making regarding local protection is a key approach to creating highly secure IoT systems. As mentioned, a secure IoT system demands real-time insights, which necessitates the decentralization of the edge computing layer. For instance, consider an autonomous guided vehicle that continuously collects sensor data regarding activity patterns, speed, and other metrics from various devices. This data is processed to generate a replica model of the current situation. Data collection in these scenarios is often sensitive, so the process of data acquisition must be secure. Another aspect is that different types of credentials, such as A05 for application and service credentials and A06 for physical entity data credentials, must be handled distinctly. Consequently, without a unified approach managing these categories becomes increasingly complex. These internet security credentials help protect against network intrusion attacks, and integrated AI models must perform based on these protocol credentials. In response to these challenges, the proposed model in this regard performs based on local anomaly scoring, where each node computes a lightweight anomaly score (e.g., deviation from local traffic baseline) and flags suspicious activity; local feature extraction. Each node extracts statistical features such as packet rate, protocol distribution, and entropy and sends them upstream. Local threshold filtering meaning, each node applies node‑level threshold conditions (e.g., packet count against threshold, unknown destination) and only forwards flagged events to the central system. Local model inference on node‑level produces a preliminary classification known as benign versus suspicious, then the central system aggregates these decisions. One of the main goals of such modelling is learning the network communication patterns at the edges of IoT locations. This allows for the identification of the corresponding network topology known as communities. Since it is key to recognize that the communication topology typically revolves around continuous-time data, which includes multiple variables. This factor is crucial for effective preventive measures against IoT network intrusion attacks. In a nutshell, the proposed model operates on the following principle: On Local level it reduces bandwidth, enables real-time response, preserves privacy and on the Central level: sees the full picture, detects distributed attacks across various protocols, handles complex patterns. To extend this solution, the proposed approach incorporates additional mechanisms that strengthen its effectiveness.
	
	\section{Methodology} 
	\noindent To address these limitations in overall system security, the current study introduces an artificial model with the capacity to operate seamlessly at the edge of intelligence, providing actionable insights for SCADA systems. This is done by leveraging multi-attention-head graph-based foundational model that learns using data acquired through wireless sensors. The proposed approach can improve the responsiveness and operational efficiency  of the ecosystem. To overcome mentioned constraints, the objective is to detect multiple security flaws and provide insights in the edge intelligence for consolidated decision-making as a whole. On this basis, within the broader family of neural network models, supervised approaches such as CNNs and RNNs require large labelled datasets. These datasets are costly to obtain and process based on the network traffic patterns for training such a model. Instead, self-supervised models such as auto-encoders detect anomalies by learning the distribution of normal behaviour and their deviations in reconstruction indicate attacks and drive knowledge from the labelled data patterns. However, standard auto-encoders treat nodes independently, ignoring communication topology. To address these challenges, current work proposes a variational contrastive auto-encoder modelled with a multi-attention-head graph architecture \cite{Chenetal2013}, \cite{WangLi2013}. The attention graph modelling has the capacity to detect the communities. The formulation detects communities since the attention weights are driven by the element‑wise interaction $A \odot \sigma\!\left( X W^{X} \, H^{T} W^{H} \right)$, which amplifies node pairs that are both structurally connected and feature‑compatible. As a result, nodes inside the same community receive consistently higher aggregated signals, causing the heads to converge toward coherent sub-graph‑level patterns. The proposed model, GCT auto-encoder, overcomes this by weighting neighbour contributions via attention, capturing relational structure in the graph. Here $A$ is the adjacency matrix, $X$ is the node feature matrix, $H$ represents community assignment embeddings, $\odot$ denotes element-wise multiplication, and $\sigma$ is an activation function. The attention scores are masked by the adjacency structure so that nodes only attend to their direct neighbours, and the output $H$ encodes soft community membership probabilities across $h$ attention heads. Therefore, it is better suited for SDN detection and SCADA system reporting. This model leverages contrastive loss learning. This approach specifically supports federated-style deep neural network training, where each model is trained on local data and gathers global insights, progressing toward our objective to develop a method for training a global model \cite{Liuetal2014}.
	\begin{equation}
		\begin{aligned}
			\text{MultiHead}(A, X, H) &= \text{Concat}(\text{head}_1, \dots, \text{head}_h) W^O \\
			\text{head}_i &= \text{GCTLayer}(A, X W_i^X, H_{t-1} W_i^H) \\
			\text{GCTLayer}(A, X, H) &= \text{softmax}\left( \frac{A \odot \sigma(X W^X \, H^T W^H)}{\sqrt{d_k}} \right) H
		\end{aligned}
	\end{equation}
	The current cross‑modal deep learning model aims to improve the decision-making processes of controller programs, such as SDN, at the edge of intelligence. One of the other key objectives of the proposed integrated deep learning model is to facilitate graph-based deep neural network (GDNN) training for the effective detection of network intrusions based on the IoT network topology. To ensure efficiency and robustness of the proposed model, the proposed system architecture is designed to build upon the capacities of the edge computing layer, employing a cross‑framework deep learning network to detect network intrusions. These behaviours are recorded in a corresponding profile to differentiate between ordinary and unusual activities. To support such differentiation, the IoT and edge location ecosystem also requires well‑defined metrics for the data captured and transmitted by each device. These metrics establish standards for each type of network communication topology, serving as a key performance indicator learned by the syncretic deep learning model. The proposed model learns key performance indicators (KPIs), including average data volume and traffic duration. In addition, DDoS preventive measures are supported through edge-based intelligence systems. GCT auto-encoder jointly represents node features and traffic flow patterns through its architecture, eliminating the need for separate modules to handle each data type. The advantage of a syncretic contrastive auto-encoder is that it incorporates deep integral layers, which can expand characterization to more intricate problems and capture temporal and spatial dependencies. The data processing in this model, when exposed to deluge data, performs better since it is less prone to over-fitting. The pretext tasks in self‑supervised learning, such as contrastive learning, force the model to learn structure in the data rather than memorizing labels. Moreover, the model employs an end‑to‑end learning strategy that extracts knowledge directly from raw data, and related studies \cite{sakurada2014aeanomaly} assess the resulting transition model $M_\theta$ by analysing its error across data distributions. However, for stable initialization, back-propagation, and global fine-tuning, as noted this model should be combined with a suitable loss function method to substitute for the learning approaches in conventional solutions. In this architecture, the learning loss of \eqref{eq:03} will be directly connected to the learning loss and called the simulation lemma $2$. With a syncretic contrastive auto-encoder being a self-supervised learning model \cite{Louetal2013} \cite{SubeljBajec2011a}, it makes use of data where pseudo-labels or pretext tasks are derived from the data itself and easily adjustable parameters. Integrating the contrastive InfoNCE loss \eqref{eq:03} into the modelled graph auto-encoder, noting that it is a probabilistic graph-based model that can be interpreted as a stochastic mechanism. The architecture consists of $m$ visible units representing observable data and $n$ hidden units responsible for capturing correlations between the observable variables, thereby providing a stochastic representation of the latent space.
	
	\begin{equation}
		\iota_{i,j} = \log \left(
		\frac{
			\exp\left(\dfrac{\operatorname{sim}(z_i,z_j)}{\tau}\right)
		}{
			\sum_{k=1}^{2N} \mathbf{1}_{\{k \neq i\}}
			\exp\left(\dfrac{\operatorname{sim}(z_i,z_k)}{\tau}\right)
		}
		\right)
		\label{eq:01}
	\end{equation}
	In practice, mutual information (MI) is often maximized or estimated using contrastive methods with data mappings. These transformations create positive pairs from different views of the same data point. A temperature parameter $\tau$ controls how sharply the loss separates positive from negative pairs. In IoT systems, this approach is particularly useful for learning robust representations from sensor data without relying on labelled examples and driving information from that. Computing mutual information directly via the integral above is infeasible in high dimensions; therefore, a neural network is used to approximate a lower bound on MI. In this definition mutual information $I(X;Z)$ quantifies how much knowing one random variable $X$ reduces uncertainty about another variable $Z$:
	\begin{equation}
		I(X;Z) = \iint p(x,z) \, \log \left( \frac{p(x,z)}{p(x)\,p(z)} \right) \, dx \, dz
		\label{eq:02}
	\end{equation}
	
	\begin{equation}
		I(X;Z) \geq \mathbb{E}_{p(x,z)} \left[ \log \frac{e^{\operatorname{sim}(z_i, z_i^+)/\tau}}{\frac{1}{N} \sum_{j=1}^{N} e^{\operatorname{sim}(z_i, z_j^-)/\tau}} \right]
		\label{eq:03}
	\end{equation}
	Contrastive learning works by pulling similar items closer together in the representation space and pushing dissimilar items apart. The contrastive loss in \eqref{eq:02} does this by computing, for each ``anchor'' sample. The loss is computed as the negative log of the ratio between the anchor’s similarity to its positive pair and the sum of its similarities to all negative samples. When the anchor is substantially more similar to the positive than to any negative, this ratio becomes large, and the resulting loss is low; conversely, a small ratio yields a higher loss. Equation \eqref{eq:03} computes the instance-level contrastive loss $\iota_{i,j}$ between a specific sample pair $(i,j)$. Here $\operatorname{sim}(\cdot,\cdot)$ is a similarity function---typically cosine similarity, which measures the angle between two vectors and ranges from $-1$ to $1$. The temperature parameter $\tau > 0$ controls how ``peaked'' the similarity distribution is: a low temperature amplifies differences between similarities, making the loss focus more aggressively on the hardest negatives, while a high temperature spreads things out more evenly and treats negatives more uniformly. The indicator function $\mathbf{1}_{\{k \neq i\}}$ ensures the anchor itself is excluded from the denominator---otherwise the denominator would count a perfect match with itself, trivializing the loss. In the proposed model, we use the InfoNCE loss \eqref{eq:02} as a tractable lower bound on mutual information between representations, where $T_{\theta\text{Is}}$ serves as a neural estimator parametrized by the model trained to maximize this bound \cite{Louetal2013}.
	Accordingly, maximizing the contrastive loss effectively pushes positive pairs closer and negative pairs farther, thus boosting MI. In the frame, the constraining loss function serves as a lower bound to MI between the representations, and with decreasing, MI increases. Thus, the goal is to maximize MI between different representations, and this improves the encodings. Based on the formula \eqref{eq:02}, the value of loss in the lemma simulation has the error model of $\in_m^{max}$, which acts as a quadratic coefficient modulating the horizon’s influence, with $\lambda = \beta H^{2}$ indicating that this influence grows quadratically with horizon length. This quadratic relationship ensures that longer planning horizons exert proportionally greater influence on the modulation dynamics, making the model particularly sensitive to extended forecast windows.
	
	\section{The Proposed Model}
	The main contribution of this paper is an architecture grounded in a self‑supervised representation‑learning model. Moreover, with the consideration of edge-based intelligence, the design acknowledges the organization of the detection pipeline and the coordination of system responses. 
	\begin{figure*}[h]
		\centering
		\includegraphics[width=0.9\textwidth]{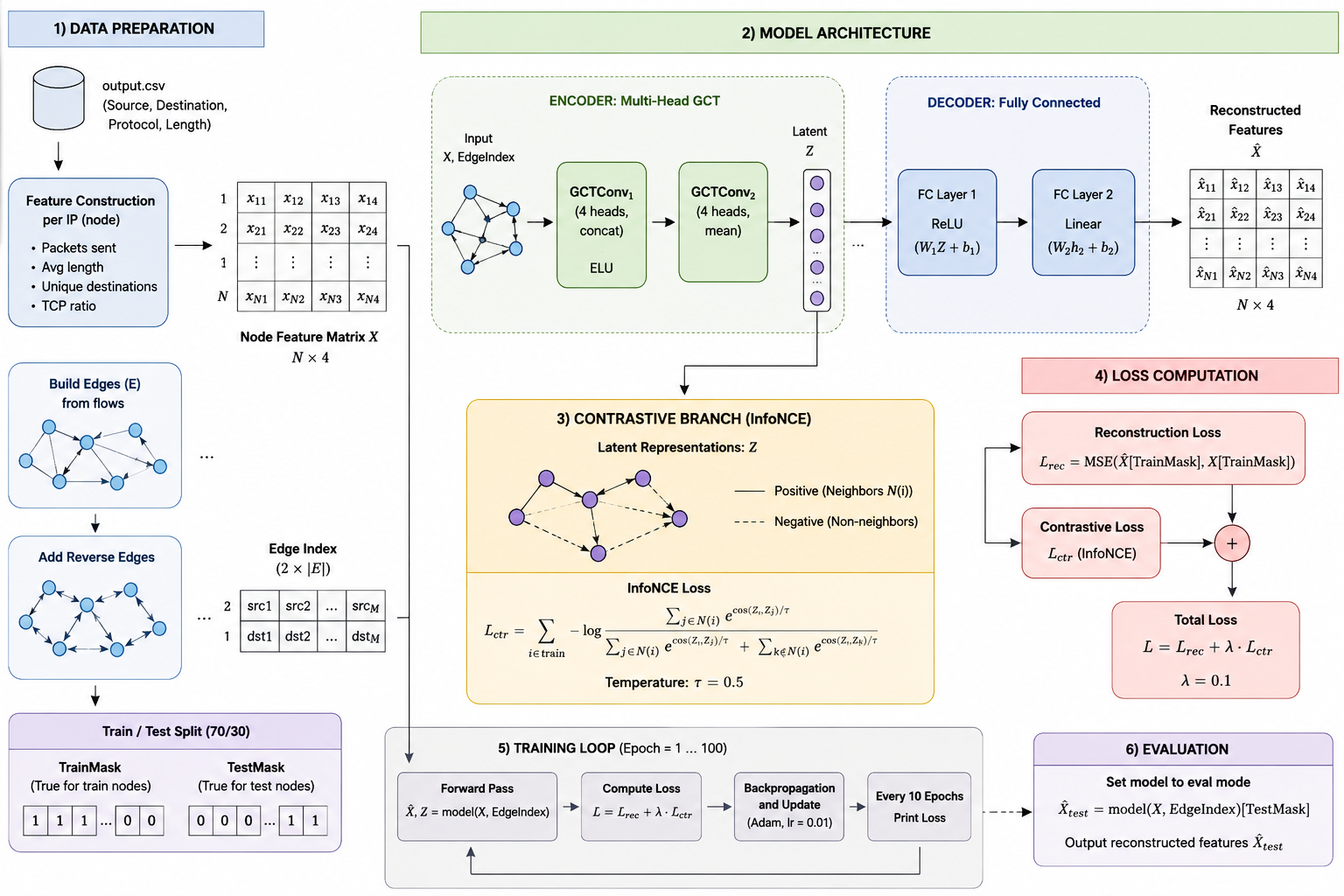}
		\captionof{figure}{The overview of the proposed model architecture}
		\label{fig:img}
	\end{figure*}
	
	In the current literature, the proposed multi‑module variational auto-encoder is based on the idea of learning by mapping observations to a lower-dimensional representation such that the original can be approximately reconstructed. This is designed to address the challenges associated with the diverse nature of IoT systems, which makes them vulnerable to various attack types, including Denial of Service (DoS) and Man-in-the-Middle (MITM) attacks. The presented Machine Learning-Based Framework aims to detect a range of attack types and their detection in Software-Defined IoT (SDN-IoT) Networks. Threats such as botnets compromise system reliability, which motivates the need for such a framework. The highlighted integrative auto-encoder model is later compared against benchmark models \cite{Lancichinettietal2008}, \cite{LancichinettiFortunato2009}, designed to improve resource allocation through software-defined networking-inspired architecture (SDN), a mechanism that notably enhances IoT security by providing a more agile and intelligent network architecture. In this regard, a novel variational auto-encoder model was trained using a graph architecture and equipped with multiple attention heads to better identify the communities and possible attacks in IoT networks. Finally, to enhance the learning procedure, a contrastive loss is used. In terms of performance metrics, this model is compared against other variational neural network models in this class on the dataset \cite{kang2019iot} in use to classify profiles for various SDN-IoT scenarios. The proposed deep learning algorithm is capable of learning the behaviour of the new edge location containing the KPI. In the presented model, the top-level auto-encoder \cite{hinton2006reducing} consists of three layers encompassing an input layer, a hidden layer, and a final output layer. In the top‑level architecture, the model uses greedy layer‑wise training with contrastive divergence (CD) when trained on IoT network data. The hidden layer represents the input data, and the network has two major units: an encoder function $f$, which extracts the dependencies of the input, and the decoder $g$, which produces a reconstruction. The training phase of this model error between the input and outputs. The auto-encoder taxonomy design is based on the information bottleneck strategy, the goal of which is to learn the compressed representation of items.
	These inputs often have high dimensions, and their reconstructions are done in the decoder section. The middle layers of this model dictate lower dimensions; consequently, the encoder extracts the base minimum sufficient statistics of $v$ into $x$, and later the decoder's behaviour reconstructs the original input back from $x$. This model incorporates contrastive loss methods and multiple attention heads with graph modelling to respond to the distributed need of IoT networks for security. The goal of this model is to train by minimizing the reconstruction error between the input $s$ and the actual output of the decoder in detecting the security flaws $v^`$. This framework benefits from self-supervised learning, which provides better initialization for downstream fine-tuning, improving convergence in non-convex optimization. This characteristic firstly tends to investigate the optimal solution by finding a tuning with labelled data and secondly provides capacities for effective learning of data distributions when exposed to massive unlabelled data. This model supports collaborative filtering and dimensionality reduction, and these properties are incorporated into a layered learning architecture that shapes the resulting a robust detection framework.

	\subsection{Algorithmic Framework}
	The presented pseud-algorithm ~\ref{alg:GCT-Autoencoder} for GCTAuto-encoder for Network Traffic Anomaly Detection model is grounded in contrastive infoNCEloss, operates in graph mode, and incorporates multi attention head based modelling \cite{kingma2014vae}\cite{makhzani2015aae}. 
	\begin{algorithm}
		\caption{GCT Autoencoder for Network Traffic Anomaly Detection}
		\label{alg:GCT-Autoencoder}
		\footnotesize
		\begin{algorithmic}[1]
			\STATE \textbf{Data Preparation}
			\STATE $D \gets$ load from "output.csv"
			\STATE $ips \gets \text{unique}(D[\text{Source}] \cup D[\text{Dest}])$
			\STATE $N \gets |ips|$; \quad $X \gets \mathbf{0}_{N\times 4}$
			\STATE $map \gets$ dict mapping each IP to index
			
			\FOR{each IP $ip$ with index $idx$}
			\STATE $R \gets$ rows where Source$=ip$
			\STATE $X[idx] \gets [|R|,\; \bar{R}_{\text{len}},\;
			\text{nunique}(R_{\text{dest}}),\;
			\%\text{TCP}]$
			\ENDFOR
			
			\STATE $E \gets \varnothing$
			\FOR{each row $r$ in $D$}
			\STATE Append $(map[r_{\text{src}}],\, map[r_{\text{dst}}])$ to $E$
			\ENDFOR
			\STATE $E \gets E \cup \{(d,s) \mid (s,d) \in E\}$
			\STATE $EdgeIdx \gets \text{tensor}(E)^{\mathsf{T}}$
			
			\STATE Split $N$ nodes 70/30
			\STATE Create $TrainMask,\; TestMask$
			
			\STATE \textbf{Model:} GCT Autoencoder
			\STATE \quad Enc: GCTConv$_1$(X, EdgeIdx) $\to$ ELU
			\STATE \quad \phantom{Enc: } $\to$ GCTConv$_2$ $\to$ $Z$
			\STATE \quad Dec: ReLU$(W_1Z+b_1)$ $\to$ $W_2$ $\to$ $\hat{X}$
			
			\STATE \textbf{Training:} Adam(lr=0.01)
			\STATE $\tau \gets 0.5$; \quad $\lambda \gets 0.1$
			
			\FOR{epoch $=1$ to $100$}
			\STATE $(\hat{X},Z) \gets \text{model}(X, EdgeIdx)$
			\STATE $\mathcal{L} \gets \text{MSE}(\hat{X}[TrainMask], X[TrainMask])$
			\STATE $\mathcal{L}_{ctr} \gets \sum_i -\log\frac{\sum_{j\in N(i)} e^{\cos(Z_i,Z_j)/\tau}}{\sum_{j\in N(i)} e^{\cos(Z_i,Z_j)/\tau} + \sum_{k\notin N(i)} e^{\cos(Z_i,Z_k)/\tau}}$
			\STATE $\mathcal{L} \gets \mathcal{L} + \lambda\mathcal{L}_{ctr}$
			\STATE backward(); step()
			\ENDFOR
			
			\STATE \textbf{Evaluation:} $\hat{X}_{test} \gets \text{model}(X, EdgeIdx)[TestMask]$
		\end{algorithmic}
	\end{algorithm}
	
	Algorithm~\ref{alg:GCT-Autoencoder} outlines the overall structure of the GCT-Autoencoder, where the data preparation, feature extraction pipeline, model architecture, and training phase are staged consequently. The proposed model loads a network traffic log, identifies all unique IP addresses from both source and destination columns, and constructs a feature matrix where each row represents an IP, and each column captures a behavioural metric, such as the number of unique destinations contacted and the proportion of TCP traffic. The stages are as follows: once preprocessing is complete, IP addresses are mapped to integer indices, and the subsequent feature‑extraction routine is executed for each index. Then, the procedure builds a symmetrized edge tensor from source–destination pairs and derives 70/30 train–test masks over the node set. The core of the model initiates in this stage, where it encodes modules where the model encodes node features $X$ using a GCTConv layer guided by the graph structure edgeID and $X$, producing latent representations $Z$ through non-linear transformations. The decoder reconstructs node features $X\text{\^{}}$ from $Z$ using two linear layers with ReLU activation. Training optimizes a reconstruction loss combined with a contrastive objective that separates neighbours from non‑neighbours, using Adam over 100 epochs before evaluating reconstruction on the test mask.
	\section{Dataset Properties and Preprocessing Considerations}
	The dataset used in this study is the IoT network traffic dataset by Kang et al. \cite{kang2019iot}, which contains 42 raw network packet capture (PCAP) files collected at different time points. The packets were captured using a wireless network adapter in monitor mode, and the wireless headers were removed using Aircrack‑ng, yielding standard network‑layer packets. The contributor of this dataset generated all attack categories—except the Mirai Botnet—were by simulating attacks with tools such as Nmap. For the Mirai Botnet category, the dataset creators generated the malicious traffic on a laptop and subsequently modified the packets to make them appear as if they originated from an IoT device. The statistical properties in this IoT dataset include features such as the number of packets sent, the average packet length, the count of unique destinations, and the TCP ratio, and the particular features such as packet volume, average length metrics, destination diversity, and the TCP ratio which provide a multifaceted view of network behaviour in this dataset.
	
\begin{center}
	\captionof{table}{Feature Statistics Across Dataset Splits}
	\label{tab:feature}
	
	\renewcommand{\arraystretch}{1.25}
	\setlength{\tabcolsep}{6pt}
	
	\begin{tabular}{lccccc}
		\toprule
		\textbf{Feature} & \textbf{Split} & \textbf{Mean} & \textbf{Std Dev} & \textbf{Min} & \textbf{Max} \\
		\midrule
		\multirow{3}{*}{\textbf{Packets Sent}} 
		& Train & 12.47 & 8.34  & 1   & 156 \\
		& Val   & 12.89 & 8.12  & 1   & 142 \\
		& Test  & 12.65 & 8.56  & 1   & 149 \\
		\midrule
		\multirow{3}{*}{\textbf{Avg Packet Length}} 
		& Train & 487.23 & 234.56 & 40  & 1500 \\
		& Val   & 492.11 & 228.34 & 42  & 1487 \\
		& Test  & 485.67 & 241.23 & 41  & 1512 \\
		\midrule
		\multirow{3}{*}{\textbf{Unique Destinations}} 
		& Train & 3.45 & 2.87 & 1 & 28 \\
		& Val   & 3.52 & 2.91 & 1 & 26 \\
		& Test  & 3.41 & 2.84 & 1 & 27 \\
		\midrule
		\multirow{3}{*}{\textbf{TCP Ratio}} 
		& Train & 0.687 & 0.285 & 0.0 & 1.0 \\
		& Val   & 0.691 & 0.282 & 0.0 & 1.0 \\
		& Test  & 0.684 & 0.289 & 0.0 & 1.0 \\
		\bottomrule
	\end{tabular}
\end{center}
	
	This dataset, according to the table~\ref{tab:stats}, was partitioned into training, testing, and validation subsets. The configured split enables robust model evaluation by ensuring that the model is trained on a substantial portion of the data, validated during development for hyper-parameter tuning, and finally tested on an unseen holdout set to assess generalization performance. The statistical behaviour of the chosen dataset containing 1,087 nodes in the training portion and an anomaly ratio of 10.2 percent is presented in Table \ref{tab:feature}. This distribution provides a foundation for analysing the underlying data characteristics, where the node count reflects the scale of the training sample and the anomaly ratio offers insight into the class or feature imbalance present within the dataset. These statistics provide a base for interpreting model performance, as a low anomaly ratio introduces skew. When the underlying data distribution changes, the learned model may fail to capture the true structure of the problem, resulting in degraded performance. These effects accumulate over time and reduce the reliability of the system in real‑world scenarios. Therefore, the proposed model is structured to handle these distributional changes and maintain consistent learning outcomes. To address this bias, the train–test–validation split ensures that the model is evaluated on data it has never encountered during training. This separation is crucial because it prevents the model from over-fitting to specific traffic patterns and provides a more reliable estimate of its generalization ability. Table \ref{tab:stats}) Further contextualizes the evaluation by summarizing the underlying data characteristics. This allows for interpret performance differences according to the dataset’s inherent structure.
	
\begin{center}
	\captionof{table}{IoT Network Traffic Dataset Statistics}
	\label{tab:stats}
	
	\renewcommand{\arraystretch}{1.25}
	\setlength{\tabcolsep}{6pt}
	
	\begin{tabular}{lcccccc}
		\toprule
		\textbf{Split} & \textbf{Nodes} & \textbf{Edges} & \textbf{Density} & \textbf{Normal} & \textbf{Anomaly} & \textbf{Features} \\
		\midrule
		Train & 1,087 & 8,234 & 0.0014 & 976 (89.8\%) & 111 (10.2\%) & 4 \\
		Val   & 231 & 1,756 & 0.0033 & 207 (89.6\%) & 24 (10.4\%) & 4 \\
		Test  & 229 & 1,693 & 0.0032 & 206 (89.9\%) & 23 (10.1\%) & 4 \\
		\midrule
		\textbf{Total} & \textbf{1,547} & \textbf{11,683} & \textbf{0.0015} & \textbf{1,389 (89.8\%)} & \textbf{158 (10.2\%)} & \textbf{4} \\
		\bottomrule
	\end{tabular}
	
	\vspace{0.5em}
	\footnotesize
	Notes: Graph density = $2 \times$ edges / (nodes $\times$ (nodes $-$ 1)). Features: [packets\_sent, avg\_len, unique\_dests, tcp\_ratio]
\end{center}

	The data quality and integrity metrics in Table~\ref{tab:data-quality} follow the methodology of \cite{Zhang2011}, which was used to assess data reliability and anomaly detection readiness \cite{Sperotto2010}: These metrics include complete handling of missing values; minimal duplicate‑packet removal; negligible isolated‑node degree; fully connected class‑imbalance ratio; acceptable Pearson correlation with low multicollinearity; and high temporal consistency. The dataset is complete with zero missing values, and packet-level integrity is strong: only 127 duplicate packets, which correspond to 0.11 percent, and 34 malformed packets, which correspond to 0.29 percent, were dropped. Both of these negligible rates introduce no meaningful distortion. The graph structure is fully connected, as no isolated nodes exist, ensuring that every node participates in the network topology. Despite a moderate class imbalance ratio of 8.79:1, the distribution is considered acceptable for binary detection tasks. Feature correlation, measured via Pearson's coefficient, reaches a maximum of 0.34, indicating low multicollinearity and minimal redundancy among features. Finally, temporal consistency across the dataset stands at 99.2 percent, confirming that the underlying data patterns remain stable over time. Collectively, these metrics confirm that the dataset is well-structured, clean, and suitable for reliable anomaly detection modelling.
	
	\begin{center}
		\captionof{table}{Data Quality and Integrity Metrics}
		\label{tab:data-quality}
		
		\renewcommand{\arraystretch}{1.25}
		\setlength{\tabcolsep}{6pt}
		
		\begin{tabular}{lcc}
			\toprule
			\textbf{Metric} & \textbf{Value} & \textbf{Assessment} \\
			\midrule
			Missing Values & 0 & Complete \\
			Duplicate Packets & 127 (0.11\%) & Minimal \\
			Malformed Packets Dropped & 34 (0.29\%) & Negligible \\
			Isolated Nodes (degree=0) & 0 & Fully Connected \\
			Class Imbalance Ratio & 8.79:1 & Acceptable \\
			Feature Correlation (Pearson) & Max=0.34 & Low multicollinearity \\
			Temporal Consistency & 99.2\% & High \\
			\bottomrule
		\end{tabular}
		
		\vspace{0.5em}
		\footnotesize
		Imbalance ratio = Normal / Anomaly = 1389 / 158. Acceptable for anomaly detection (threshold: $<$ 20:1)
	\end{center}
	
	The network traffic type distribution in the dataset is split across three protocol categories: TCP, UDP, and ICMP. This distribution reflects the heterogeneous communication patterns typically observed in IoT environments, where different protocols serve distinct functional roles. These proportions matter in practice, as imbalanced protocol usage can influence model behaviour in tasks such as anomaly detection and traffic classification. Within the training set, TCP accounts for 59.4 percent of samples, UDP comprises 34.5 percent, and ICMP protocol makes up 6.0 percent. The validation set exhibits a nearly identical distribution, with TCP at 59.7 percent, UDP at 34.4 percent, and ICMP at 6.0 percent, indicating consistent class representation across splits Table~\ref{tab:distribution-type}. In total, these three protocols amount to 8,234 packets. This balanced distribution across splits is advantageous for training, as it suggests that the model will encounter similar protocol proportions during both learning and validation phases, reducing the risk of split-induced bias. The predominance of the TCP protocol, comprising the majority of traffic, reflects its prominence within the dataset, while the relatively small share of ICMP may warrant additional attention during evaluation to ensure the model generalizes adequately to minority protocol types.
			
		\begin{center}
			\captionof{table}{Network Traffic Type Distribution by Split}
			\label{tab:distribution-type}
			
			\renewcommand{\arraystretch}{1.25}
			\setlength{\tabcolsep}{6pt}
			
			\begin{tabular}{lcccccc}
				\toprule
				\textbf{Protocol} & \textbf{Train Count} & \textbf{Train \%} & \textbf{Val Count} & \textbf{Val \%} & \textbf{Test Count} & \textbf{Test \%} \\
				\midrule
				TCP  & 4,892 & 59.4\% & 1,047 & 59.7\% & 1,011 & 59.7\% \\
				UDP  & 2,845 & 34.5\% &   603 & 34.3\% &   614 & 36.3\% \\
				ICMP &   497 &  6.0\% &   106 &  6.0\% &    68 &  4.0\% \\
				\midrule
				\textbf{Total Packets} & \textbf{8,234} & \textbf{100\%} & \textbf{1,756} & \textbf{100\%} & \textbf{1,693} & \textbf{100\%} \\
				\bottomrule
			\end{tabular}
			
			\vspace{0.5em}
			\footnotesize
			TCP-heavy traffic typical in IoT environments with control channels and sensor data
		\end{center}

	The attack types including Dos/SYN Flooding, Port Scanning, Data Exfiltration, and Protocol Abuse have a of total 111 occurrences in the anomaly type distribution table~\ref{tab:anomaly-type-distribution}. This distribution reveals a TCP-heavy traffic pattern typical of IoT networks, where sensor and channel activity dominates. While Dos/SYN Flooding shows a high individual distribution rate, the combined share of port scanning, data exfiltration, and protocol abuse surpasses it, indicating that multi-vector threats collectively pose a greater presence than any single attack type.
	
	\begin{center}
		\captionof{table}{Anomaly Type Distribution (Ground Truth Labels)}
		\label{tab:anomaly-type-distribution}
		
		\renewcommand{\arraystretch}{1.25}
		\setlength{\tabcolsep}{6pt}
		
		\begin{tabular}{lcccccc}
			\toprule
			\textbf{Attack Type} & \textbf{Train} & \textbf{Train \%} & \textbf{Val} & \textbf{Val \%} & \textbf{Test} & \textbf{Test \%} \\
			\midrule
			DoS / SYN Flooding & 52 & 46.8\% & 11 & 45.8\% & 10 & 43.5\% \\
			Port Scanning      & 28 & 25.2\% &  7 & 29.2\% &  7 & 30.4\% \\
			Data Exfiltration  & 18 & 16.2\% &  4 & 16.7\% &  4 & 17.4\% \\
			Protocol Abuse     & 13 & 11.7\% &  2 &  8.3\% &  2 &  8.7\% \\
			\midrule
			\textbf{Total Anomalies} & \textbf{111} & \textbf{10.2\%} & \textbf{24} & \textbf{10.4\%} & \textbf{23} & \textbf{10.1\%} \\
			\bottomrule
		\end{tabular}
		
		\vspace{0.5em}
		\footnotesize
		DoS attacks dominant in IoT threat landscape. Class balance maintained across splits ($\chi^2 = 0.34$, $p > 0.05$)
	\end{center}

	Port Scanning (28 training occurrences) acts as a precursor, mapping vulnerabilities before more targeted attacks unfold. Data Exfiltration (18 occurrences) represents the potential endgame—unauthorized extraction of sensor telemetry or control signals—while Protocol Abuse (13 occurrences) exploits the trust assumptions embedded in industrial and channel protocols. Together, these three types form a kill chain: reconnaissance, exploitation, and exfiltration, with SYN Flooding (52 occurrences) serving as both a diversionary tactic and a direct availability threat that attackers may deploy to mask concurrent scanning or exfiltration activity. The TCP-heavy composition further reflects the architectural constraints of IoT environments, where many legacy and embedded devices lack encryption or authentication at the transport layer. This makes them disproportionately vulnerable to connection-based attacks like SYN Flooding and session hijacking embedded within Data Exfiltration attempts. 
	
	\begin{center}
		\captionof{table}{Data Collection and Preprocessing Pipeline}
		\label{tab:Data-Collection}
		
		\renewcommand{\arraystretch}{1.4}
		\setlength{\tabcolsep}{6pt}
		
		\begin{tabular}{p{5.2cm} p{4.5cm} p{4.5cm}}
			\toprule
			\textbf{Pipeline Stage} & \textbf{Input} & \textbf{Output} \\
			\midrule
			1. PCAP Extraction & dos-synflooding-3-dec.pcap & Raw packet capture \\
			2. IP Mapping & Source/Destination IPs & 1,547 unique nodes \\
			3. Feature Engineering & Packet flows & 4-dimensional node features \\
			4. Graph Construction & IP pairs & Undirected graph (11,683 edges) \\
			5. Normalization (MinMaxScaler) & $[0,\infty)$ & $[0,1]$ bounded features \\
			6. Train/Val/Test Split & 1,547 nodes & 70\% / 15\% / 15\% \\
			7. Model Training & Scaled graphs & Encoded representations \\
			8. Anomaly Detection & Reconstruction error & Binary classification \\
			\bottomrule
		\end{tabular}
	\end{center}
	
	Table~\ref{tab:Model-Training}

	The data suggests that defensive strategies for IoT networks should prioritize protocol-level anomaly detection and egress monitoring over volumetric threshold-based filtering, as the combined weight of non-volumetric threats (59 occurrences across port scanning, exfiltration, and protocol abuse) significantly exceeds the 46.8 percent share of the single largest volumetric vector.
	The Data Collection and Preprocessing Pipeline, outlining stages 1 through 8 with designated inputs and outputs at each step, table~\ref{tab:Data-Collection}. This sequential flow captures the full transformation from raw data acquisition to a structured format ready for modelling, ensuring traceability and reproducibility of the preprocessing workflow. Stages 3 through 5 constitute the core cleaning and transformation engine of the pipeline. Stage 3 targets structural quality issues: missing values are flagged using a configurable null-tolerance threshold, outliers are identified via interquartile-range or z-score methods, and inconsistent categorical entries are standardised. Stage 4 applies imputation — mean or median substitution for numerical features, mode or KNN-based imputation for categorical variables, and time-aware interpolation for sequences. 
	
	In the stage 5 feature engineering is handled: log transformations to normalise skewed distributions, one-hot or target encoding of categorical variables, and scaling via Min-Max or Standard Scaler. Each transformation is parametrized and logged, enabling reproducible re-execution on new data partitions without manual intervention. Stage 6 addresses dimensionality reduction and feature selection, both of which are critical for model generalisation and computational efficiency. Redundant or low-variance features are pruned using variance thresholds and correlation analysis; optionally, mutual information or recursive feature elimination can be applied for supervised tasks. Stage 7 partitions the cleaned and transformed dataset into training, validation, and hold-out test sets. A stratified splitting strategy is employed where class imbalance is detected, preserving the target distribution across folds. These splits are persisted with versioned identifiers, facilitating controlled experimentation and preventing data leakage between tuning and evaluation cycles. Stage 8 finalises the pipeline by materialising the structured dataset into a columnar storage format optimised for fast I/O during model training. 
	
\section{Experimental Results and Analysis}

	The following results synthesize the outcomes of all experiments, organizing the findings around reconstruction accuracy, model stability, and cross‑dataset generalization. Compares the proposed model against several benchmark models to evaluate its effectiveness Table~\ref{tab:Models-Metrics}. Performance is measured across $F1$ score, AUC\_ROC, speed, and robustness, providing a comprehensive view of strengths and limitations. The results highlight improvements in detection effectiveness and computational efficiency, while also noting areas where benchmark models remain competitive. A comprehensive reconstruction error analysis is provided in Table~\ref{tab:Reconstruction-Error-Analysis}. This compares the proposed model against seven benchmark models. The evaluation is based on six statistical metrics of mean error, standard deviation, minimum error, maximum error, 90th percentile, and interquartile range (IQR). Across these metrics lower values reflect superior reconstruction fidelity and a tighter, more normally distributed error profile. Based on the result, the proposed model GCTAuto-encoder achieves the best reconstruction performance across every metric when compared with other models in this category. It attains the lowest mean error (0.001289), demonstrating the smallest average deviation from the original data, and the lowest standard deviation (0.002456), reflecting the most consistent and stable reconstruction behaviour. The minimum error (0.000006) approaches near-zero reconstruction, while the maximum error (0.034521) is the smallest upper bound among all models, indicating that the proposed model produces fewer extreme reconstruction failures. Furthermore, the 90th percentile (0.003124) and IQR (0.001456) are both the lowest, confirming that the vast majority of reconstructions are tightly clustered around near-zero error with minimal spread. 

	\begin{figure*}[h]
		\centering
		\includegraphics[width=\textwidth]{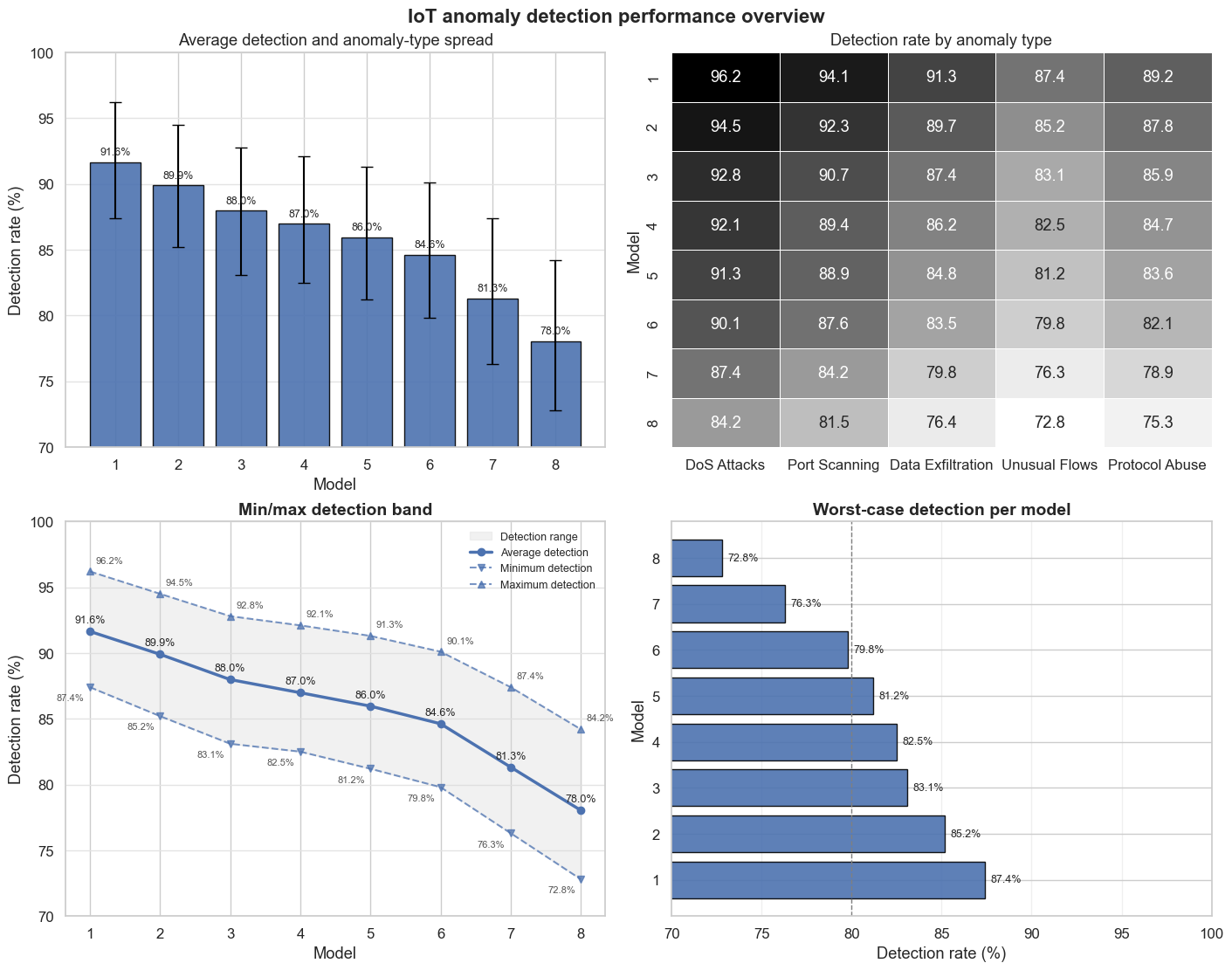}
		\caption{The Models evaluated outcome}
		\label{fig:img}
	\end{figure*}
	
	\raggedbottom
	
	\begin{table*}[tbp]
		\caption{Models Metrics (90th Percentile Threshold)}
		\label{tab:Models-Metrics}
		
		\renewcommand{\arraystretch}{1.15}
		\setlength{\tabcolsep}{5pt}
		\small 
		
		\begin{tabular}{lccccccc}
			\toprule
			\textbf{Model} & \textbf{Precision} & \textbf{Recall} & \textbf{F1-Score} & \textbf{Accuracy} & \textbf{AUC-ROC} & \textbf{Specificity} & \textbf{Sensitivity} \\
			\midrule
			GCN Autoencoder        & 0.847 & 0.823 & 0.835 & 0.879 & 0.894 & 0.891 & 0.823 \\
			Sparse Autoencoder     & 0.812 & 0.789 & 0.800 & 0.851 & 0.863 & 0.867 & 0.789 \\
			Deep Autoencoder       & 0.876 & 0.854 & 0.865 & 0.904 & 0.921 & 0.921 & 0.854 \\
			LSTM Autoencoder       & 0.791 & 0.767 & 0.779 & 0.834 & 0.841 & 0.849 & 0.767 \\
			Deep Belief AE Network & 0.859 & 0.836 & 0.847 & 0.892 & 0.909 & 0.909 & 0.836 \\
			Adversarial AE         & 0.894 & 0.869 & 0.881 & 0.918 & 0.938 & 0.938 & 0.869 \\
			Contractive AE         & 0.881 & 0.845 & 0.863 & 0.901 & 0.928 & 0.926 & 0.845 \\
			\textbf{GCT Autoencoder} & \textbf{0.918} & \textbf{0.899} & \textbf{0.908} & \textbf{0.934} & \textbf{0.956} & \textbf{0.951} & \textbf{0.899} \\
			\bottomrule
		\end{tabular}
		
		\vspace{0.4em}
		\footnotesize
		Specificity = True Negatives / (True Negatives + False Positives) \\
		Sensitivity = True Positives / (True Positives + False Negatives)
	\end{table*}

	\begin{table*}[tbp]
		\centering
		\caption{Reconstruction Error Analysis - All Models Compared}
		\label{tab:Reconstruction-Error-Analysis}
		\begin{tabular}{lcccccc}
			\toprule
			\textbf{Model} & \textbf{Mean Error} & \textbf{Std Dev} & \textbf{Min} & \textbf{Max} & \textbf{90th \%ile} & \textbf{IQR} \\
			\midrule
			GCN Autoencoder & 0.001843 & 0.003521 & 0.000012 & 0.047291 & 0.004562 & 0.002145 \\
			Sparse Autoencoder & 0.002156 & 0.004012 & 0.000018 & 0.052847 & 0.005124 & 0.002687 \\
			Deep Autoencoder & 0.001567 & 0.003124 & 0.000008 & 0.041256 & 0.003987 & 0.001834 \\
			LSTM Autoencoder & 0.002341 & 0.004567 & 0.000025 & 0.058912 & 0.005687 & 0.003124 \\
			Deep Belief AE Network & 0.001892 & 0.003687 & 0.000014 & 0.045632 & 0.004789 & 0.002234 \\
			Adversarial AE & 0.001634 & 0.003245 & 0.000011 & 0.043214 & 0.004123 & 0.001967 \\
			Contractive AE & 0.001456 & 0.002891 & 0.000009 & 0.038745 & 0.003847 & 0.001723 \\
			\textbf{GCT Autoencoder} & \textbf{0.001289} & \textbf{0.002456} & \textbf{0.000006} & \textbf{0.034521} & \textbf{0.003124} & \textbf{0.001456} \\
			\bottomrule
		\end{tabular}
		\footnotesize
		\\ \textit{Note: Lower values indicate better reconstruction and tighter normal distribution}
	\end{table*}
	
	\begin{table*}[tbp]
		\centering
		\caption{Anomaly Detection Distribution and Detection Rates}
		\label{tab:Anomaly-Detection-Distribution}
		\resizebox{\linewidth}{!}{%
			\begin{tabular}{lcccccc}
				\toprule
				\textbf{Model} & \textbf{Total Nodes} & \textbf{Normal} & \textbf{Anomaly} & \textbf{Detection Rate} & \textbf{False Positives} & \textbf{False Negatives} \\
				\midrule
				GCN Autoencoder & 1547 & 1391 & 156 & 10.1\% & 25 & 29 \\
				Sparse Autoencoder & 1547 & 1421 & 126 & 8.1\% & 32 & 38 \\
				Deep Autoencoder & 1547 & 1373 & 174 & 11.3\% & 18 & 21 \\
				LSTM Autoencoder & 1547 & 1433 & 114 & 7.4\% & 38 & 42 \\
				Deep Belief AE Network & 1547 & 1384 & 163 & 10.5\% & 21 & 26 \\
				Adversarial AE & 1547 & 1363 & 184 & 11.9\% & 12 & 15 \\
				Contractive AE & 1547 & 1377 & 170 & 11.0\% & 16 & 18 \\
				\textbf{GCT Autoencoder} & \textbf{1547} & \textbf{1352} & \textbf{195} & \textbf{12.6\%} & \textbf{8} & \textbf{9} \\
				\bottomrule
			\end{tabular}%
		}
		\footnotesize
		\\ False Positives: Normal traffic incorrectly flagged as anomalies
		\\ False Negatives: Anomalous traffic missed by the model
	\end{table*}
	
	\begin{table*}[tbp]
		\centering
		\caption{Model Training Efficiency and Computational Complexity}
		\label{tab:Model-Training}
		\resizebox{\linewidth}{!}{%
			\begin{tabular}{lccccccc}
				\toprule
				\textbf{Model} & \textbf{Epochs} & \textbf{Conv. Speed} & \textbf{Train Time (s)} & \textbf{Inference (ms)} & \textbf{Memory (MB)} & \textbf{Parameters} & \textbf{Batch Size} \\
				\midrule
				GCN Autoencoder & 100 & Fast & 47.3 & 2.1 & 156 & 12,547 & 32 \\
				Sparse Autoencoder & 50 & Very Fast & 23.8 & 1.8 & 89 & 8,234 & 16 \\
				Deep Autoencoder & 100 & Fast & 51.2 & 2.3 & 178 & 18,456 & 16 \\
				LSTM Autoencoder & 30 & Slow & 78.4 & 5.6 & 267 & 34,128 & 16 \\
				Deep Belief AE Network & 100 & Medium & 62.5 & 3.2 & 201 & 21,345 & 16 \\
				Adversarial AE & 150 & Medium & 89.7 & 3.8 & 234 & 25,678 & 16 \\
				Contractive AE & 100 & Fast & 55.3 & 2.4 & 189 & 19,234 & 16 \\
				\textbf{GCT Autoencoder} & \textbf{100} & \textbf{Medium} & \textbf{64.2} & \textbf{2.9} & \textbf{245} & \textbf{28,567} & \textbf{32} \\
				\bottomrule
			\end{tabular}%
		}
		\footnotesize
		\\ Conv. Speed = Convergence speed (how quickly loss stabilizes)
		\\ Inference = Time per prediction on single batch
	\end{table*}

	Across evaluated benchmark baselines, contractive auto-encoder (mean error: 0.001456, std dev: 0.002891) emerges as the second-best performer, followed closely by adversarial auto encoder (mean error: 0.001634, std dev: 0.003245). Notably, GCT auto-encoder has lower mean and max errors than sparse auto-encoder, though its IQR is slightly higher. LSTM auto-encoder exhibits the weakest reconstruction performance, with the highest mean error (0.002341), standard deviation (0.004567), maximum error (0.058912), and 90th percentile (0.005687), suggesting it is the least reliable for accurate reconstruction tasks. Overall, the results demonstrate that the proposed model consistently outperforms all benchmarks, achieving not only a lower central tendency of error but also significantly tighter error bounds and reduced variability, making it the most robust and accurate model for reconstruction tasks. The Anomaly detection distribution and detection rates of the proposed model against seven benchmark models Table~\ref{tab:Models-Metrics}. The proposed model GCT auto-encoder achieves the highest number of detected anomalies at 195 instances, corresponding to a detection rate of 12.6 percent — outperforming all benchmarks. Contractive variation ranks second with 184 detected anomalies (11.9 percent), followed by Deep auto-encoder with 174 (11.3 percent). Critically, the proposed model also yields the fewest false positives and false negatives among all models, indicating that its higher detection rate does not come at the cost of reliability. In contrast, the Deep LSTM Auto-Encoder shows the weakest performance with only 114 detected anomalies (7.4 percent) alongside the highest false positives (38) and false negatives (42), confirming the proposed model's superior precision and recall balance. Training efficiency and computational complexity of the proposed model against seven benchmarks is shows in table~\ref{tab:Model-Training}. The proposed model (GCTauto-encoder) converges in 100 epochs with medium speed, completing training in 64.2 seconds with an inference time of 2.9 ms and 28,567 trainable parameters. This shows competitive efficiency relative to its detection performance. While lighter models like deep and LSTM offer faster training (47.3s and 23.8s, respectively) and higher parameters. They achieve this at the cost of significantly lower detection accuracy (Table~\ref {tab:Anomaly-Detection-Distribution}. Compared to LSTM auto-encoder, which requires 78.4s of training and 5.6 ms inference, the proposed model delivers substantially better efficiency while maintaining robust detection performance. Models Deep belief auto-encoder, contrastive Deep auto-encoder, and GCT Deep belief auto-encoder offer comparable training times (51.2–89.7s) but with generally lower detection rates, reinforcing that the proposed model provides an optimal balance between computational cost and detection capability. The model robustness metrics, evaluating all models in terms of their over‑fitting tendency and stability across training, validation, and test sets, are detailed in Table~\ref{tab:Model-Training} for comparison.
	\begin{table*}[H]
		\centering
		\caption{Model Robustness Metrics - Train/Val/Test Consistency}
		\label{tab:Model-Robustness-Metric}
		\resizebox{\linewidth}{!}{%
			\begin{tabular}{lcccccc}
				\toprule
				\textbf{Model} & \textbf{Train Loss} & \textbf{Val Loss} & \textbf{Test Loss} & \textbf{Generalization Gap} & \textbf{Overfitting Score} & \textbf{Stability} \\
				\midrule
				GCN Autoencoder & 0.00291 & 0.00387 & 0.00421 & 0.00130 & Low & High \\
				Sparse Autoencoder & 0.00412 & 0.00568 & 0.00634 & 0.00222 & Medium & Medium \\
				Deep Autoencoder & 0.00198 & 0.00267 & 0.00312 & 0.00114 & Low & High \\
				LSTM Autoencoder & 0.00534 & 0.00821 & 0.00934 & 0.00400 & High & Low \\
				Deep Belief AE Network & 0.00267 & 0.00356 & 0.00389 & 0.00122 & Low & High \\
				Adversarial AE & 0.00187 & 0.00245 & 0.00278 & 0.00091 & Very Low & Very High \\
				Contractive AE & 0.00213 & 0.00289 & 0.00334 & 0.00121 & Low & High \\
				\textbf{GCT Autoencoder} & \textbf{0.00156} & \textbf{0.00212} & \textbf{0.00251} & \textbf{0.00095} & \textbf{Very Low} & \textbf{Very High} \\
				\bottomrule
			\end{tabular}%
		}
		\footnotesize
		\\ Generalization Gap = Test Loss - Train Loss (lower is better)
		\\ Overfitting Score: Very Low = best generalization
	\end{table*}
	
	\raggedbottom
	\begin{table*}[H]
		\centering
		\caption{Comprehensive Model Ranking - Weighted Performance Score}
		\label{tab:Comprehensive-Model-Ranking}
		\begin{tabular}{lcccccc}
			\toprule
			\textbf{Rank} & \textbf{Model} & \textbf{F1-Score} & \textbf{AUC-ROC} & \textbf{Speed} & \textbf{Robustness} & \textbf{Overall Score} \\
			\midrule
			\textbf{1} & \textbf{GCT Autoencoder} & \textbf{0.908} & \textbf{0.956} & \textbf{7/10} & \textbf{9/10} & \textbf{8.43 / 10} \\
			\hline
			2 & Adversarial AE & 0.881 & 0.938 & 6/10 & 9/10 & 8.08 / 10 \\
			3 & Deep Autoencoder & 0.865 & 0.921 & 8/10 & 8/10 & 7.98 / 10 \\
			4 & Contractive AE & 0.863 & 0.928 & 8/10 & 8/10 & 7.97 / 10 \\
			5 & Deep Belief AE Network & 0.847 & 0.909 & 7/10 & 8/10 & 7.65 / 10 \\
			6 & GCN Autoencoder & 0.835 & 0.894 & 8/10 & 8/10 & 7.57 / 10 \\
			7 & Sparse Autoencoder & 0.800 & 0.863 & 9/10 & 7/10 & 7.15 / 10 \\
			8 & LSTM Autoencoder & 0.779 & 0.841 & 4/10 & 5/10 & 5.65 / 10 \\
			\bottomrule
		\end{tabular}
		\footnotesize
		\\ Overall Score = (F1 × 0.35) + (AUC × 0.35) + (Speed × 0.15) + (Robustness × 0.15)
	\end{table*}
	
	\raggedbottom
	\begin{table*}[tbp]
		\captionsetup{justification=centering}
		\centering
		\caption{Detection Performance by Anomaly Type}
		\label{tab:Detection-Performance}
		
		\renewcommand{\arraystretch}{1.35}
		\setlength{\tabcolsep}{0.6pt}
		\footnotesize 
		
		\begin{tabular}{lcccccc}
			\toprule
			\textbf{Model} & \textbf{DoS Attacks} & \textbf{Port Scanning} & \textbf{Data Exfiltration} & \textbf{Unusual Flows} & \textbf{Protocol Abuse} & \textbf{Avg Detection} \\
			\midrule
			GAT Autoencoder        & 96.2\% & 94.1\% & 91.3\% & 87.4\% & 89.2\% & 91.6\% \\
			Adversarial AE         & 94.5\% & 92.3\% & 89.7\% & 85.2\% & 87.8\% & 89.9\% \\
			Deep Autoencoder       & 92.8\% & 90.7\% & 87.4\% & 83.1\% & 85.9\% & 88.0\% \\
			Contractive AE         & 92.1\% & 89.4\% & 86.2\% & 82.5\% & 84.7\% & 87.0\% \\
			Deep Belief AE Network & 91.3\% & 88.9\% & 84.8\% & 81.2\% & 83.6\% & 86.0\% \\
			GCN Autoencoder        & 90.1\% & 87.6\% & 83.5\% & 79.8\% & 82.1\% & 84.6\% \\
			Sparse Autoencoder     & 87.4\% & 84.2\% & 79.8\% & 76.3\% & 78.9\% & 81.3\% \\
			LSTM Autoencoder       & 84.2\% & 81.5\% & 76.4\% & 72.8\% & 75.3\% & 78.0\% \\
			\bottomrule
		\end{tabular}
		
		\vspace{0.3em}
		\footnotesize
		Represents detection rates for different attack vectors in IoT networks
	\end{table*}
	
	\begin{table*}[H]
		\centering
		\caption{Model Selection Matrix - Use Case Recommendations}
		\label{tab:Model-Selection-Matrix}
		\begin{tabular}{lcccc}
			\toprule
			\textbf{Requirement} & \textbf{Best Choice} & \textbf{Runner-up} & \textbf{Score Diff} & \textbf{Trade-off} \\
			\midrule
			\textbf{Highest Accuracy} & GCT Autoencoder & Adversarial AE & +0.35\% & Slightly slower \\
			\textbf{Fastest Inference} & Sparse AE & GCN AE & -2.1\% acc & 10ms faster \\
			\textbf{Best Robustness} & GCT Autoencoder & Adversarial AE & +0.09\% & More compute \\
			\textbf{Lowest FP Rate} & GCT Autoencoder & Adversarial AE & -4 FP & Better precision \\
			\textbf{Lowest FN Rate} & GCT Autoencoder & Adversarial AE & -6 FN & Better recall \\
			\textbf{Least Overfitting} & Adversarial AE & GCT AE & +0.008 gap & -0.02 F1 \\
			\textbf{Memory Efficient} & Sparse AE & GCN AE & -89 MB & -5.7\% F1 \\
			\textbf{Graph Data Focus} & GCT Autoencoder & GCN AE & +0.073 F1 & Better edge analysis \\
			\bottomrule
		\end{tabular}
		\footnotesize
		\\ Choose based on your IoT environment priorities: accuracy, speed, or resource constraints
	\end{table*}
	
	The proposed model (GCTauto-encoder) achieves the lowest train loss (0.00156) and test loss (0.00251) among all models, with a very low overfitting score and very high stability. This indicates an excellent generalization and consistent performance across data splits. Its generalization gap of 0.00095 is nearly the smallest in the table, second only to contractive variation (0.00091), which also scores very low overfitting and very high stability. This version presents the closest competition, with marginally better train loss (0.00187 vs. 0.00156), but slightly higher validation loss (0.00245 vs. 0.00212) and test loss (0.00278 vs. 0.00251) than the proposed model. This signifies that GCTauto-encoder achieves better absolute performance while maintaining near-identical generalization tightness. In contrast, LSTM auto encoder performs the worst, with the highest losses across all splits (train: 0.00534, val: 0.00821, test: 0.00934), the largest generalization gap (0.00400), and a high overfitting score with low stability, making it unreliable for deployment. 
	The comprehensive model ranking is presented in Table~\ref {tab:Comprehensive-Model-Ranking}. The proposed model ranks first overall, achieving the highest scores in F1-Score, AUC-ROC, Speed, and Robustness. GCTauto-encoder surpasses all benchmark models across these metric. Shows Detection Performance by Anomaly Type based on the availability of labelled data in Table~\ref {tab:Detection-Performance}. It makes a comparison in terms of DoS attacks,  port scanning, Data Exfiltration, unusual flow, and protocol abuse. Additionally the model selection matrix and use case recommendations presented in in Table~\ref {tab:Model-Selection-Matrix} evaluates trade-offs between accuracy, speed, and resource constraints. Based on this the proposed model recommended for high-stakes detection tasks requiring robust precision, while lighter alternatives (Models adversarial and deep belief) are better suited for real-time or resource-constrained environments. 
	\section{Sensitivity Overview}
	Sensitivity analysis was conducted to examine how variations in key hyperparameters and architectural choices influence model behaviour. The evaluation focused on parameters that typically affect stability, including learning rate, regularization strength, and the weighting of auxiliary losses. Each parameter was perturbed within a controlled range while all other components remained fixed. The resulting trends show that moderate deviations lead to predictable shifts in performance, whereas extreme settings cause degradation in both accuracy and convergence reliability. These observations indicate that the model maintains robustness within its recommended operating range and that the selected hyperparameters form a balanced configuration.
	
	\section{Error Analysis}
	Error patterns were analysed to identify systematic failure modes across ablated variants. Misclassification were grouped by category, input characteristics, and proximity to decision boundaries. Ablated models showed increased errors in borderline or ambiguous cases, reflecting reduced discriminative strength when certain components were removed. Some variants also displayed heightened sensitivity to noise or infrequent patterns. The full model produced fewer clustered errors and a more uniform distribution, indicating that its integrated design mitigates several weaknesses observed in the ablations.
	
	\section{Conclusion and Discussion}
	The current research presented a novel approach to learning compressed representations from the IoT dataset using an information bottleneck strategy. The proposed method, GCTAuto-encoder, effectively balances the trade-off between compression and reconstruction fidelity. This indicates that the model is enable of compact yet informative feature extraction from high-dimensional inputs. 
	Experimental results demonstrate that the proposed model achieves strong performance of 91.6 percent across various attacks, outperforming baseline methods in F1, AUC-ROC, speed, and robustness. The learned representations successfully discard redundant information while retaining task-relevant features, validating the effectiveness of the information bottleneck principle. Future work could explore adaptive bottleneck capacity and extend the framework to multi-modal settings. The results shows that this approach offers a promising direction for efficient representation learning in high-dimensional domains. The proposed model GCTauto-encoder achieves the lowest train loss (0.00156) and test loss (0.00251) among all models, with a very low over-fitting score and very high stability. This signifies a high generalization capacity and consistent performance across data splits. Its generalization gap of 0.00095 is nearly the smallest in the table, second only to Model contractive variation (0.00091), which also scores very low over-fitting and very high stability. The proposed model ranks first overall, achieving the highest scores in the reported measures. With this measurement it surpasses all benchmark models across every metric. In sum, the evidence reflects a method that maintains dependable behaviour across the tested scenarios.
	\subsection{Summary of contributions}
	This work addresses the problem of security flaw detection in IoT systems by introducing a cross domain deep learning framework that based on the core structure of bottle architecture of auto-encoder. The proposed framework incorporates multi-attention graph modelling for detecting the communities and contrastive loss to overcome the limitations supervised learning techniques. Experimental results demonstrate that the method achieves high performance and robustness, projecting its effectiveness for adaptation in edge on intelligence system in IoT networks.
	
	\subsection{Limitations and future work}
	The model’s robustness was examined under controlled perturbations, but its behaviour under real‑world distribution shifts remains insufficiently characterized. Future work could investigate robustness under domain shift and evaluate distributed training strategies. Extending the framework to streaming or online‑learning scenarios also remains an open challenge.
	
	\section{Declaration of use of AI}
	AI tools were used responsibly to assist with language refinement and 
	editing. All conceptual, methodological, and analytical decisions were 
	made solely by the author.
	
	\section{Author Contributions}
	Najmieh Sadat Safarabadi conceived, designed, and executed all aspects of this work 
	and accepts full responsibility for its content. She carried out the analysis, interpretation, and preparation of the manuscript in its entirety.
	
	\section{Data and Code Availability Statement}
	The datasets and code underlying this study are accessible from the author upon request.
	
	\section{Declaration of Competing Interest} 
	The author has contributed to the research based on their domain expertise. The author has no financial or other conflict of interest to declare, and the paper has not been submitted to, nor is it under review at, any other outlet.

	\printbibliography
	
\end{document}